\documentclass[12pt]{article}
\usepackage{amsmath,amssymb,graphicx,mathrsfs,hyperref}

\newcommand{\be}{\begin{equation}}
\newcommand{\ee}{\end{equation}}
\newcommand{\bea}{\begin{eqnarray}}
\newcommand{\eea}{\end{eqnarray}}

\def\({\left(} \def\){\right)}

\begin{document}

\title{\vspace{-1.8in}
\vspace{0.3cm} {Lovelock gravity is equivalent to Einstein gravity coupled to form fields}}
\author{\large Ram Brustein${}^{(1)}$,  A.J.M. Medved${}^{(2)}$ \\
 \hspace{-1.5in} \vbox{
 \begin{flushleft}
  $^{\textrm{\normalsize
(1)\ Department of Physics, Ben-Gurion University,
    Beer-Sheva 84105, Israel}}$
$^{\textrm{\normalsize (2)  Department of Physics \& Electronics, Rhodes University,
  Grahamstown 6140, South Africa }}$
 \\ \small \hspace{1.7in}
    ramyb@bgu.ac.il,\  j.medved@ru.ac.za
\end{flushleft}
}}
\date{}
\maketitle

 \begin{abstract}

Lovelock gravity is a class of higher-derivative gravitational theories whose linearized equations of motion  have no more than two time derivatives. Here, it is shown that any Lovelock theory can be effectively described as Einstein gravity coupled to a p-form gauge field. This extends the known example of an {$f(R)$} theory of gravity, which can be described as Einstein gravity coupled to a scalar field.

\end{abstract}
\newpage

\section{Introduction}

The Lovelock class of gravity theories are the unique set of higher-derivative extensions to  Einstein gravity  with  two-derivative field equations \cite{Lovelock}.
When one quantizes small perturbations about a fixed background,
this two-derivative limit on the equations of motion is practically equivalent to the unitarity of the theory. We will use the two notions interchangeably.

The interaction terms in a theory of  Lovelock gravity  are constrained by the dimensionality $D$ of the  spacetime.  A Lovelock term with $2k$ derivatives is purely
topological  when $\;2k=D\;$ and identically vanishing when $\;2k>D\;$. So that, formally, $\;k\leq k_{\rm max}\leq\frac{D-1}{2}\;$. For a more general and recent perspective on this issue,
see \cite{cavepeeps}.

Over the years, modified theories of gravity differing from the Lovelock class have been proposed.
However, even if unitarity is maintained, all of these include non-trivial modifications to Einstein gravity such as additional gravitons modes (either in the way
of massive gravitons or bi-metric theories) and/or spacetimes
of dimension less than four. (For a short sample of more recent work, see \cite{mod1}-\cite{mod8}.) Such proposals are interesting in their own right but, unlike Lovelock
gravity,  present serious  phenomenological issues and have no obvious connection to the realm of string theory.

That unitary gravity is equivalent to Lovelock's theories may appear to be a powerful assertion to some \cite{robbed} and
a natural one to others \cite{Zwi}. Regardless, we now intend  to ``raise the ante'' and further argue that unitary gravity is essentially equivalent to Einstein's theory only. The basis for this new claim is that any theory of Lovelock gravity can be effectively described as Einstein gravity coupled to a matter
field, much in the same way that any model of $f(R)$ gravity can be described as Einstein gravity coupled to a scalar field \cite{f_R}. For $D$-dimensional  Lovelock gravity, the role
of the scalar is played by a $(D-3)$-form gauge field.

We proceed to elaborate on this picture.
After some preliminary considerations, we discuss the cases of $f(R)$ gravity (as a warmup), Gauss--Bonnet gravity and, then, Lovelock gravity of arbitrary order. The paper concludes with a brief summary.

Recently  in \cite{Blumenhagen}, the association between Lovelock gravity and form fields was noted and some possible relations to string theory were proposed.

\section{Lovelock gravity}

Lovelock gravity has the following Lagrangian:
\be
L_{LL}\;=\;\sum_{k=0}^{k_{\rm max}} \lambda_{k} L_{k} \;,
\ee
where $L_{k}$ contains terms with  $k$ Riemann tensors  contracted together, the $\lambda$'s are dimensionful coupling constants and the sum runs up to $\;k_{\rm max}\leq \frac{D-1}{2}\;$. In this set-up, $L_0$ is the cosmological constant $\Lambda$, $\;L_1=L_E\;$ is the Ricci scalar and $L_2$ is the Gauss--Bonnet extension~(\ref{GBext}) which  will be encountered shortly. The other terms are defined in Eq.~(\ref{mess}) below.

Let us define the tensor
$\;
{\cal X}^{abcd} \equiv \frac {\delta L}{\delta {\cal R}_{abcd}}\;
$,
which has the symmetry properties of
the Riemann tensor ${\cal R}_{abcd}$. Lovelock theories can be defined by the identity \cite{Lovelock}
\be
\nabla_a {\cal X}^{abcd} \;=\;0 \;,
\label{love}
\ee
valid on- or off-shell and  satisfied by each the Lovelock terms
separately.

The field equation for a Lovelock theory takes the form
\cite{Wald,BGHM}
$\;
{\cal X}^{abc}_{\;\;\;\;\;\;p}{\cal R}_{abcq}-\frac{1}{2}g_{pq} L =  \frac{1}{2}T_{pq}\;,
$
with $T_{ab}$ being the stress-tensor for the matter fields.
For Einstein gravity, $\;L_E={\cal R}^{ab}_{\;\;\;\;ab}\;$ (we work with
$\;16\pi G=1\;$ units throughout)  and
$\;
{\cal X}_E^{abcd} =  \frac{1}{2}\left[g^{ac}g^{bd}-g^{ad}g^{bc}\right]\;,
$
which leads to the standard Einstein equation.

A note on conventions: ``Calligraphic letters'' are used to denote  four-index tensors.
So that, for instance, an $R$ means the Ricci scalar whereas
an ${\cal R}$ is meant as shorthand for ${\cal R}^{abcd}$.
 A ``dot product'' will signify a  four-fold contraction,
 $\;{\cal A}\cdot{\cal B}={\cal A}^{abcd}{\cal B}_{abcd}\;$.

The explicit form of the Lovelock term $L_{k}$  is given by
(with the usual symmetrization factor absorbed into $\lambda_k$) \cite{Lovelock}
\be
L_{k}\;=\; \delta^{a_1b_1\ldots a_kb_k}_{c_1d_1 \ldots c_k d_k}\;{\cal R}^{c_1 d_1}_{\;\;\;\;\;\;a_1b_1}
\cdots{\cal R}^{c_k d_k}_{\;\;\;\;\;\;a_kb_k}\;,
\label{mess}
\ee
where the generalized alternating tensor $\;\delta^{\cdots}_{\cdots}\;$ is fully anti-symmetric
in both its upper and lower indices. For example, $\;\delta^{pq}_{rs }=
\left[\delta^{p}_{\;r}\delta^{q}_{\;s}- \delta^{p}_{\;s}\delta^{q}_{\;r}\right]$.  Hence, we find that
$\;
{\cal X}_k^{abcd} =  \frac {\delta L_k}{\delta {\cal R}_{abcd}}\;
$
goes as
\be
\left({\cal X}_{k}\right)^{pq}_{\;\;\;\;\;rs}\;=\; k\;\delta^{pq a_2 b_2\ldots a_kb_k}_{rsc_2d_2
\ldots c_k d_k}\;{\cal R}^{c_2 d_2}_{\;\;\;\;\;\;a_2b_2}
\cdots{\cal R}^{c_k d_k}_{\;\;\;\;\;\;a_kb_k}\;.
\label{messX}
\ee

The ${\cal X}_{k}$'s satisfy the first and second Bianchi identities. The first Bianchi identity $\;\left({\cal X}_{k}\right)^{pq}_{\;\;\;\;\;rs}+\left({\cal X}_{k}\right)^{ps}_{\;\;\;\;\;qr}+\left({\cal X}_{k}\right)^{pr}_{\;\;\;\;\;sq}=0\;$ is trivially satisfied because of the Riemannian symmetry of  $\;\delta^{pq \cdots}_{rs\cdots }$.

The second Bianchi identity $\;\nabla_m\left({\cal X}_{k}\right)^{pq}_{\;\;\;\;\;rs}+\nabla_s\left({\cal X}_{k}\right)^{pq}_{\;\;\;\;\;mr}+\nabla_r\left({\cal X}_{k}\right)^{pq}_{\;\;\;\;\;sm}=0\;$ follows from the structure of ${\cal X}_{k}$ and
the fact that the Riemann tensor satisfies the second Bianchi identity. Differentiating ${\cal X}_k$, while keeping in mind the
product rule and the interchangeability
of the Riemann tensors, we have
\be
\nabla_m \left({\cal X}_{k}\right)^{pq}_{\;\;\;\;\;rs}\;=\;
k(k-1)\;\delta^{pq a_2 b_2\ldots a_kb_k}_{rsc_2d_2
\ldots c_k d_k}\;\nabla_m\left[{\cal R}^{c_2 d_2}_{\;\;\;\;\;\;a_2b_2}\right]
{\cal R}^{c_3 d_3}_{\;\;\;\;\;\;a_3 b_3}
\cdots{\cal R}^{c_k d_k}_{\;\;\;\;\;\;a_kb_k}\;.
\label{B2a}
\ee
Now, permuting the first two pairs of upper indices
and using the Riemannian symmetries of the $\delta$-symbol,
we obtain
\bea
\nabla_m \left({\cal X}_{k}\right)^{pq}_{\;\;\;\;\;rs} & = &\cdots
\left[\delta^{a_2}_{\ r}\delta^{b_2}_{\ s}- \delta^{a_2}_{\ s}\delta^{b_2}_{\ r}\right]
\nabla_m \left[{\cal R}^{c_2 d_2}_{\;\;\;\;\;\;a_2b_2}\right]\cdots
\nonumber \\
& = & \cdots \nabla_m\left[R^{c_2d_2}_{\;\;\;\;\;\;rs}\right]\cdots\;.
\label{B2b}
\eea
The second Bianchi identity then follows.

\section{$f(R)$ gravity revisited}

Let us first consider how $\; L(R)=R^n\;$ models
are equivalent to Einstein--scalar theories \cite{f_R},
but  from our  novel perspective.
For these models,
$\;n L(R)= {\cal X}\cdot {\cal R}\;$ such that
$\;{\cal X}^{abcd}=  n R^{n-1}\ {\cal X}_E^{abcd}\;$.

We propose the following effective description:
\bea
n\widetilde{L}({\cal R},{\cal S},{\cal Y})\; &=&\; {\cal S}\cdot{\cal Y}
 \;+ \; {\cal Y}\cdot\left({\cal R}-{\cal S}\right)
 \;+ \; {\cal S}\cdot\left({\cal X}-{\cal Y}\right)\; \cr
&=& {\cal Y}\cdot\left({\cal R}-{\cal S}\right)
\;+ \; {\cal S}\cdot{\cal X} \cr
&=& {\cal Y}\cdot{\cal R}
\;+ \; {\cal S}\cdot\left({\cal X}-{\cal Y}\right)\;,
\label{efflag}
\eea
with ${\cal S}^{abcd}$ and  ${\cal Y}^{abcd}$ serving as auxiliary tensor fields.

Varying $\widetilde{L}$ by
${\cal S}$, one obtains $\;{\cal Y}={\cal X}\;$. Similarly, $\;{\cal S}={\cal R}\;$ is obtained by varying $\widetilde{L}_n$
with respect to ${\cal Y}$.
Substituting  these relations into $\widetilde{L}$,  we find that, when on-shell,
$\;\widetilde{L}=L\;$  and can therefore be viewed  as an equivalent description of the same theory.

Next, we provide the auxiliary fields with  physically motivated identities, while respecting the tensorial properties of their on-shell equivalents. To begin, ${\cal Y}$ can be expressed in terms of a scalar
field $\phi$ times a tensor; say,
\be
{\cal Y}^{abcd}\;=\; \phi {\cal X}_E^{abcd}\;.
\ee
Then, just like in the standard  procedure, $\phi$ becomes a dynamical field that is equivalent to $f^\prime(R)$ on shell ($\prime$ denotes a derivative with respect to the argument).

As for ${\cal S}$, let us make the choice
\be
{\cal S}^{abcd}\;=\; \frac{2\psi}{D(D-1)} {\cal X}_E^{abcd}\;,
\ee
where  $\;\psi=\psi(\phi)\;$ is a scalar functional  and the
 ``normalization'' factor has been chosen to ensure that
the on-shell condition $\;{\cal S}={\cal R}\;$ translates into
  $\;\psi=R\;$.

The effective  Lagrangian can now be reformulated as
\bea
n\widetilde{L} \;&=&\; \phi R +\psi(\phi)\left(n R^{n-1}-\phi \right)\;\cr
\;&=&\; \phi R +\psi(\phi)\left(n [\psi(\phi)]^{n-1}-\phi \right)\;\cr
\;&=&\; \phi R -V(\phi) \;
.
\eea
The ``potential'' $V(\phi)$ is given by
$\;V\left[\phi,\psi(\phi)\right]= [\phi-f^\prime(\psi)]\psi\;$. Its minimum is determined by the equation
$\;V^\prime(\phi)=0\;$ and occurs at $\;\phi_m=n R^{n-1}\;$, as can be seen from the equation $\;{\cal Y}={\cal X}\;$.
Even though the scalar is constrained on-shell to be at the minimum of the potential, $\phi$ is still fully dynamical and, additionally,
$\phi_m$ and $\;\psi(\phi_m)=R\;$ are not necessarily
spacetime constants. This all follows from the gravitational field equation, which goes as
\hbox{$\;
\nabla_p \nabla_q \phi -\phi R_{pq}
=\frac{1}{2}g_{pq}\left[2\nabla^a\nabla_a \phi -\phi R+ V(\phi)\right]\;.
$}

And so what we end up with is the well-known, expected  result that $f(R)$ gravity is dynamically equivalent
to Einstein gravity coupled to a scalar field.

\section{Gauss--Bonnet gravity}

The simplest non-trivial Lovelock extension of Einstein gravity is  Gauss--Bonnet (GB) gravity. Its Lagrangian is given by
$\;
L_{GB} = L_1 + L_0  +  \lambda_2 L_{2}\;
$
for which
$\;
{\cal X}_{GB}^{abcd} =  {\cal X}^{abcd}_E + \lambda_2{\cal X}^{abcd}_{2}\;$.
Recall that $\;L_1 = L_E$\;, $\;L_0=\Lambda\;$  and, from Eq.~(\ref{mess}),
\bea
\lambda_2 L_{2} & = & \lambda_2\left[{\cal R}^{abcd}{\cal R}_{abcd} -4 R^{ab}R_{cd}+ R^2\right]\;.
\label{GBext}
\eea
The value of the dimensionful coupling constant $\lambda_2$ is  irrelevant to the current treatment.

Via Eq.~(\ref{messX}),
\be
{\cal X}^{abcd}_{2}\;=\;
 2\left[{\cal R}^{abcd}-g^{ac}R^{bd}-g^{bd}R^{ac}
+g^{ad}R^{bc}+g^{bc}R^{ad}+ R{\cal X}_E^{abcd}\right]\;.
\label{XGB}
\ee
To be dynamical, the GB theory requires $\;D\geq 5\;$. Then,
$
\;2L_{2}={\cal R}\cdot {\cal X}_{2}\;.
$

We focus on the GB terms $L_{2 }$, ${\cal X}_{2 }$ and closely follow the discussion of $f(R)$ gravity until arriving at the step of identifying the auxiliary tensors.   For  ${\cal S}$ (and, likewise, for ${\cal Y}$), what is required is matter  that can carry four indices and respect the basic symmetries of the Riemann tensor. We are thus driven to the choice of   $p$-form gauge fields $B^{[a_1 a_2 \cdots a_p]}$ and their totally anti-symmetrized $(p+1)$-form field-strength tensors
$\;
H^{\left[a_1 a_2 \cdots a_{p+1}\right]}=\nabla^{\left[a_1\right.}B^{\left. a_2 a_3 \cdots a_{p+1} \right]}\;,
\label{hdef}
$
and so
\bea
{\cal S}^{ab}_{\ \ cd} &=& H^{\left[e_1\cdots e_{n} ab\right]}H_{\left[e_1\cdots e_{n}cd\right]}\;,
\label{calS}
\eea
where $\;n=p-1\;$. Anti-symmetrization of indices will be implied from now on.

The $H$'s are identically vanishing unless $\;p\leq D-1\;$. To allow for four different indices on an ${\cal S}$,
the condition  $\;p\leq D-3\;$ is further required.
Even when off-shell, ${\cal S}$ satisfies the basic symmetry properties of the Riemann tensor;
for instance, $\;{\cal S}^{bacd} = -{\cal S}^{abcd}\;$,
$\;{\cal S}^{cdab} = +{\cal S}^{abcd}\;$.

It can be also be  shown that, even  when off-shell,
${\cal S}$  satisfies the first Bianchi identity.
To this end, let us use {\em vierbein} formalism to write
$
{\cal S}^{ab}_{\ \ cd} \;=\; e_{i}^{\ a} e_{j}^{\ b} e_{\ c}^{k} e_{\ d}^{l}
\; H^{i j e_1\cdots e_{n}}H_{k l e_1\cdots e_{n}}\;
$
or
\bea
{\cal S}^{ab}_{\ \ c  d}  & = & \left[A^{\ k}_{i}\right]^{a}_{ \  c}
\left[A^{\ l}_{j}\right]^{b}_{ \ d}
\; H^{i j e_1\cdots e_{n}}H_{k l e_1\cdots e_{n}} \nonumber \\
 & = & \frac{1}{2}\Big(\left[A^{\ k}_{i}\right]^{a}_{ \  c}
\left[A^{\ l}_{j}\right]^{b}_{ \ d}  -
\left[A^{\ k}_{i}\right]^{a}_{ \  d}
\left[A^{\ l}_{j}\right]^{b}_{\ c}
\Big)
 H^{i j e_1\cdots e_{n}}H_{k l e_1\cdots e_{n}}\;,
\eea
where $\;\left[A^{\ j}_{i}\right]^{a}_{ \  b}
\equiv  e_{i}^{\ a} e_{\ b}^{j}\;$
and the last identity follows from the basic Riemannian structure
of ${\cal S}$. Given  the form  $\;{\cal S}^{ab}_{\ \ c  d}\propto A^{a}_{\ c}A^{b}_{\ d}-A^{a}_{\ d}A^{b}_{\ c}\;$ for a tensor ${\cal S}$, the first Bianchi identity follows.

To obtain the GB version of ${\cal Y}$, we  apply the ansatz for ${\cal S}$ in Eq.~(\ref{calS}) to rewrite
Eq.~(\ref{messX}) as
\be
{\cal Y}_{2\ \ cd}^{\;ab} \;=\; 2\;\delta^{ab\; a_2 b_2}_{cd \;c_2d_2}\;{\cal S}^{c_2 d_2}_{\;\;\;\;\;\;a_2b_2} \;.
\label{defY2}
\ee
The explicit result in terms of $H$'s can also be obtained by replacing all the Riemann tensors in Eq.~(\ref{XGB}) with the corresponding expressions for ${\cal S}$,
\be
{\cal Y}_2^{abcd} \;=\;  2\left[H^{e_1\cdots e_{n} ab}H_{e_1\cdots e_{n}}^{\;\;\;\;\;\;\;\;\;cd}
-g^{ac}H^{e_1\cdots e_{p} b}H_{e_1\cdots e_{p}}^{\;\;\;\;\;\;\;\;\;d}
-g^{bd}H^{e_1\cdots e_{p} a}H_{e_1\cdots e_{p}}^{\;\;\;\;\;\;\;\;\;c}
\right. \nonumber
\ee
\be
\left. \;+\;
g^{ad}H^{e_1\cdots e_{p} b}H_{e_1\cdots e_{p}}^{\;\;\;\;\;\;\;\;\;c}
+g^{bc}H^{e_1\cdots e_{p} a}H_{e_1\cdots e_{p}}^{\;\;\;\;\;\;\;\;\;d}
+H^2
{\cal X}^{abcd}_E\right]\;,
\label{calY}
\ee
where  $\;H^2= H^{e_1\cdots e_{p+1}}H_{e_1\cdots e_{p+1}}\;$.
Notice that, by its definition, ${\cal Y}_2$ automatically satisfies
the first Bianchi identity.

We will assume that there are no local sources for
 the field-strength tensor ({\it i.e.}, no branes).
Then, in direct analogy to standard electromagnetism, this tensor  must  have a
vanishing divergence,
\be
\nabla_a H^{ae_1\cdots e_p} \;=\; 0 \;,
\label{divH}
\ee
and  satisfy a  Bianchi-like identity,
\be
\nabla_a H_{bc}^{\;\;\;\;\;e_1\cdots e_n} + \nabla_b H_{ca}^{\;\;\;\;\;e_1\cdots e_n} +  \nabla_c H_{ab}^{\;\;\;\;\;e_1\cdots e_n}
\;=\; 0 \;.
\label{bianH}
\ee

It turns out that the latter is enough to establish that both ${\cal S}$ and ${\cal Y}_2$ satisfy the second Bianchi identity
even off-shell. For ${\cal S}$, this is true because of the Riemannian
``double-exchange'' symmetry $\;{\cal S}^{abcd}={\cal S}^{cdab}$\;,
meaning that
\bea
\nabla_e {\cal S}_{abcd} & = & \nabla_e \left[H^{e_1\cdots e_{n}}_{\;\;\;\;\;\;\;\;\;\;\; ab} \; H_{e_1\cdots e_{n}cd}\right]
\nonumber \\
&  = & 2\left[\nabla_e\; H^{e_1\cdots e_{n}}_{\;\;\;\;\;\;\;\;\;\;\; ab}\right]\; H_{e_1\cdots e_{n}cd}
 \; = \; 2 H^{e_1\cdots e_{n}}_{\;\;\;\;\;\;\;\;\;\;\; ab}\nabla_e \;H_{e_1\cdots e_{n}cd}
\;,
\eea
and the second Bianchi identity  follows from Eq.~(\ref{bianH})
(at least) for the sets $e,a,b$ and $e,c,d$.

One might be concerned about cases in which
the set of permuted indices  starts out  on different $H$'s.  It is, however, a simple matter to use Riemannian and field-strength
(anti-) symmetry  properties to manipulate these onto the same $H$.

Since the second Bianchi identity is true for ${\cal S}$, it is  also true for ${\cal Y}_2$; this, by direct analogy with our previous argument that any ${\cal X}_k$ satisfies the second Bianchi identity
given that ${\cal R}$ does ({\em cf}, Eqs.~(\ref{B2a}-\ref{B2b})).

The GB version of the effective Lagrangian~(\ref{efflag}) can now be put in the form
\bea
\widetilde{L}_{GB} &=&
L_1+L_0  -\frac{1}{2(p+1)}  H^2 + \lambda_2\widetilde{L}_2
\nonumber \\
&=&  R +\Lambda -\frac{1}{2(p+1)}  H^2 +
\lambda_2\Big[
\frac{1}{2}{{\cal Y}_2(H)}\cdot \bigl({\cal R} - {{\cal S}(H)}\bigr) +
{{\cal S}(H)} \cdot{\cal X}_{2}
\Big] \cr
&=& R +\Lambda - \frac{1}{2(p+1)} H^2 +
\lambda_2\Big[ \frac{1}{2}{{\cal Y}_2(H)}\cdot{\cal R} +\frac{1}{2}{{\cal S}(H)}\cdot \bigl( {\cal X}_{2}-{{\cal Y}_2(H)}\bigr) \Big]
\;,
\nonumber \\
&&
\label{GBeqn}
\eea
where ${{\cal S}(H)}$ is given by Eq.~(\ref{calS}) and ${{\cal Y}_2(H)}$, by Eq.~(\ref{calY}).
The kinetic term $H^2$ has been included for completeness.

The equivalence principle is violated by the interaction terms ${\cal Y}_2\cdot {\cal R}$ and ${\cal S}\cdot{\cal X}_2$, as these non-trivially couple the Riemann curvature to the field-strength tensor.
But  our formulation makes it clear that the violation of the equivalence principle can be attributed to the gauge fields coupling with the Einstein graviton rather than  an exotic form of gravitation, in exact analogy with  the case of $f(R)$ gravity.

\subsection{Equations of motion}

We would now like to understand how unitarity --- a maximum of two time derivatives in the linearized equations of motion --- is maintained for the effective theory. Of course, the two-derivative constraint on these equations
is assured to hold on-shell, as this is when  $\widetilde{L}_{GB}$
and $L_{GB}$ are describing equivalent theories.

Let us begin with the field equation for the gauge field. Varying the effective action, we obtain the expression
\bea
\frac{\delta \widetilde L_{2}}{\delta B^{q_1\cdots q_{p} }}
  = & - &\frac{1}{(p+1)}\Biggl[(-1)^p(p-1)\nabla_c
\left({\cal X}_{2\; q_1q_2ab}H_{q_3\cdots q_{p}}^{\;\;\;\;\;\;\;\;\;cab}\right)
-2\nabla^c\Biggl({\cal X}_{2\; q_1 cab} H_{q_2\cdots q_{p}}^{\;\;\;\;\;\;\;\;\;ab}\Biggr) \cr
\;\;\;\;\;\;\;\;\;\;\hspace{-.5in} &-&(-1)^{p}(p-1)\nabla_c\left( {{\cal Y}}_{2\; q_1q_2ab}
H_{q_3\cdots q_{p}}^{\;\;\;\;\;\;\;\;\;cab}\right)
+2\nabla^c\Biggl({{\cal Y}}_{2\; q_1 cab} H_{q_2\cdots q_{p}}^{\;\;\;\;\;\;\;\;\;ab}\Biggr) \Biggr]\;.
\nonumber \\
& &
\label{gaugeom}
\eea

One might be concerned by the presence of multi-derivative terms; however,
it turns out that this variation is identically vanishing.
This outcome follows from
the Lovelock identity~(\ref{love}), the vanishing
divergence of the field strengths~(\ref{divH}), and the realization
that
both ${\cal X}_2$ and ${\cal Y}_2$ are Riemannian tensors.

To understand how all this works, let us start with  the first term on the right-hand side of Eq.~(\ref{gaugeom}).
After imposing Eq.~(\ref{divH}), we have (with some indices suppressed for clarity)
\be
\nabla_c\left(H_{\cdots}^{\;\;\;cab} {\cal X}_{qrab}\right)\;=\; H_{\cdots}^{\;\;\;cab}\nabla_c {\cal X}_{qrab}\;.
\label{use2B}
\ee
Now, since $\;H_{\cdots}^{\;\;\;cab}=H_{\cdots}^{\;\;\;abc}=H_{\cdots}^{\;\;\;bca}\;$,
 this term can be recast into
\be
H_{\cdots}^{\;\;\;cab}\nabla_c {\cal X}_{qrab}\;=\; \frac{1}{3}\left[H_{\cdots}^{\;\;\;cab}\nabla_c {\cal X}_{qrab}
+H_{\cdots}^{\;\;\;cab}\nabla_a {\cal X}_{qrbc}+H_{\cdots}^{\;\;\;cab}\nabla_b {\cal X}_{qrca}\right]\;,
\ee
which vanishes by virtue of the second Bianchi identity. The same argument can be used to establish that the third term on the right is also vanishing.

The second term on the right
of Eq.~(\ref{gaugeom}) can similarly be shown to  vanish.
Here, we start by using  Eq.~(\ref{love}) to rewrite this term as
\be
\nabla^c\left({\cal X}_{q_1 cab}H_{\cdots\; q_p}^{\;\;\;\;\;\;\;ab}\right) \; = \;
{\cal X}_{q_1 cab}\nabla^c H_{\cdots\; q_p}^{\;\;\;\;\;\;\;ab}\;.
\ee
Now consider that
\bea
\nabla^c H_{\cdots\; q_p}^{\;\;\;\;\;\;\;\;ab} &=&
\nabla_{q_p}H^{\;\;\;\;\;cab}_{\cdots}
\eea
because covariant derivatives commute when acting on a $B$ and, since
 $\;H_{\cdots}^{\;\;\;cab}=H_{\cdots}^{\;\;\;abc}= H_{\cdots}^{\;\;\;bca}\;$,
\be
\nabla^c\left({\cal X}_{q_1 cab}H_{\cdots\; q_p}^{\;\;\;\;\;\;\;ab}\right)
\;=\;
\frac{1}{3}\left[{\cal X}_{q_1abc}+
{\cal X}_{q_1bca}+{\cal X}_{q_1cab}\right]
\nabla_{q_p} H_{\cdots}^{\;\;\;\;abc}\;,
\ee
which vanishes via the first Bianchi identity.

The fourth term in Eq.~(\ref{gaugeom}) vanishes in the same way as the second does
except that, in this case,
one applies Eq.~(\ref{divH}) to move ${\cal Y}_2$ outside of the derivative.

We can establish unitarity for the linearized field equation for gravity by showing that  its associated  ${\cal X}$ satisfies the
Lovelock identity~(\ref{love}).   The variation of  $\widetilde{L}_{GB}$ with respect
to a Riemann tensor yields
\be
{\cal X}^{abcd}_{\widetilde{L}_{GB}} \;=\; {\cal X}^{abcd}_E
+\frac{\lambda_2}{2}{{\cal Y}}^{abcd}
+\frac{\lambda_2}{2}{{\cal S}}_{pqrs} \frac{\delta {\cal X}_{2}^{pqrs}}{\delta {\cal R}_{abcd}}\;.
\label{issue}
\ee

Identity~(\ref{love}) is then satisfied, as all  three terms on the right-hand side have a vanishing divergence.
This is evident for the first term via Eq.~(\ref{love})
and  the second term by way of Eq.~(\ref{divH}). The divergence of the third term vanishes due to the following argument:
As ${\cal S}$ is functionally independent of the Riemann tensor, we can express the third term as
\be
{{\cal S}}_{pqrs} \frac{\delta {\cal X}_{2}^{pqrs}}{\delta {\cal R}_{abcd}}
\;=\;
\frac{\left(\delta {\cal S}\cdot {\cal X}_2\right) }{\delta {\cal R}_{abcd}}\;.
\ee
Next, applying Eq.~(\ref{messX}),
\be
{\cal S}\cdot{\cal X}_2\;=\; 2\;\delta^{pqa_2b_2}_{rsc_2d_2}
\; {\cal S}^{rs}_{\;\;\;\;\;\;pq}\;{\cal R}^{c_2 d_2}_{\;\;\;\;\;\;a_2b_2}\;,
\ee
which leads to
\bea
{{\cal S}}_{pqrs} \frac{\delta {\cal X}_{2}^{pqrs}}{\delta {\cal R}_{ab}^{\;\;\;\;\;cd}}
& = & 2 \;\delta^{pqab}_{rscd} \;
{\cal S}^{rs}_{\;\;\;\;\;\;pq}
\nonumber
\\
& = & {\cal Y}^{ab}_{\;\;\;\;\;\;cd}\;,
\eea
with the latter equality
following from Eq.~(\ref{defY2}). Hence, this third term
is identical to  the second, and so  unitarity has been established.

\section{Lovelock gravity of arbitrary order}

For an arbitrary-order term in the Lovelock expansion, things work pretty much the same as for the GB case.
The number of $H$'s in the interaction term will increase with increasing $k$, but there are no conceptual differences. Indeed, the generalized version of the effective
Lagrangian~(\ref{GBeqn}) takes the form
$\;
\widetilde{L}_{LL}\;=\; R + \Lambda -\frac{1}{2(p+1)} H^2 + \sum\limits_{k=2}^{k^{\rm max}}\lambda_k \widetilde{L}_k\;,
$
with
$\;k\widetilde{L}_k\;=\;{{\cal Y}_k(H)}\cdot{\cal R}
+{{\cal S}(H)}\cdot{\cal X}_{k}
-{{\cal S}(H)}\cdot {{\cal Y}_k(H)} \;
$
as for the earlier-studied models.
The function ${{\cal Y}_k(H)}$ now goes as  $H^{2k-2}$ and its specific structure is determined by the structure of ${\cal X}_{k}$.

\subsection{Equations of motion}

The verification of  unitarity  follows along similar  lines to the GB case, leading to the same basic results.
For instance, the $k^{{\rm th}}$-order Lovelock term leads to a field equation of the form
\bea
0 &=&\frac{\partial \widetilde L_{k}}{\partial B^{q_1\cdots q_{p} }} \nonumber \\
 &=&  -\frac{2}{k(p+1)}\Bigg[(-1)^p (p-1)\nabla_c
\left({\cal X}_{k\; q_1q_2ab}H_{q_3\cdots q_{p}}^{\;\;\;\;\;\;\;\;\;cab}\right)
-2\nabla^c\left({\cal X}_{k\;q_1 cab} H_{q_2\cdots q_{p}}^{\;\;\;\;\;\;\;\;\;ab}\right)\Bigg. \nonumber
\eea
\be
+(-1)^p(p-1)(k-1)\nabla_c\left(
{\cal Z}_{k\; q_1q_2ab}H_{q_3\cdots q_{p}}^{\;\;\;\;\;\;\;\;\;cab}\right)
-2(k-1)\nabla_c\left({\cal Z}_{k\; q_1 cab}
H_{q_2\cdots q_{p}}^{\;\;\;\;\;\;\;\;\;ab}\right)
\nonumber\ee
\be
\Bigg. -(-1)^{p}(p-1)k\nabla_c\left({\cal Y}_{k\; q_1q_2ab}
 H_{q_3\cdots q_{p}}^{\;\;\;\;\;\;\;\;\;cab}\right)
+2k\nabla^c\left({\cal Y}_{k\; q_1 cab} H_{q_2\cdots q_{p}}^{\;\;\;\;\;\;\;\;\;ab} \right)\Bigg]\;,\nonumber
\ee
\be
\ee
where $\;
{\cal Z}_{k\; abcd}
\equiv \frac{\partial \left({\cal Y}_k\cdot {\cal R}
\right)}{\partial{\cal S}^{abcd}}\;.
$
We find  that, once again, the equation is automatically satisfied given
a vanishing divergence for $H$ plus the Riemannian symmetries
of ${\cal S}$, ${\cal Y}_2$  and ${\cal X}_2$.

The only real subtlety of the generic Lovelock analysis  might be in verifying that the last term  in the generalized version of Eq.~(\ref{issue}),
\be
\frac{1}{\lambda_k}{\cal X}^{abcd}_{\widetilde{L}_k} \;=\;
\frac{1}{k}{{\cal Y}_k}^{abcd}
+\frac{1}{k}{{\cal S}_{pqrs}} \frac{\delta {\cal X}_{k}^{pqrs}}{\delta {\cal R}_{abcd}}\;,
\label{issue2}
\ee
satisfies the Lovelock identity~(\ref{love}). That
$\;
\nabla_a \frac{{\cal S}_{pqrs}\delta{\cal X}_{k}^{pqrs}}{\delta {\cal R}_{abcd}}\;=\;0\;
$
or, equivalently,
\be
\frac{\delta^2 \left({{\cal S}} \cdot {\cal X}_{k}\right)}{\delta {\cal R}_{efgh}
\delta {\cal R}_{abcd}}\nabla_a{\cal R}_{efgh} \;+\;
\frac{\delta^2 \left({{\cal S}} \cdot {\cal X}_{k}\right)}{\delta {\cal S}_{efgh}
\delta {\cal R}_{abcd}}\nabla_a{\cal S}_{efgh}
\;=0\;\;.
\label{rerun}
\ee

For this purpose, we  again call upon the
explicit form of a Lovelock term~(\ref{mess}).
Suitably modified, this is
\be
\frac{1}{k}{{\cal S}}\cdot {\cal X}_k \;=\;
\delta^{a_1b_1\ldots a_kb_k}_{c_1d_1 \ldots c_k d_k}\;{\cal R}^{c_1 d_1}_{\;\;\;\;\;\;a_1b_1}\cdots
{\cal S}^{c_j d_j}_{\;\;\;\;\;\;a_j b_j}\cdots
{\cal R}^{c_k d_k}_{\;\;\;\;\;\;a_kb_k}\;,
\label{modmess}
\ee
where the ellipses indicate Riemann tensors only.

Now twice varying Eq.~(\ref{modmess}) with
the appropriate tensors, we have
\be
\frac{1}{k}\frac{\delta^2 \left({{\cal S}}\cdot
{\cal X}_k\right)}{\delta{\cal R}_{ef}^{\;\;\;\;gh}\partial{\cal R}_{ab}^{\;\;\;\;cd}} \;=\;
(k-1)(k-2)\;\delta^{ab e f a_3b_3\ldots  a_{k}b_{k}}_{cdgh c_3d_3 \ldots c_{k} d_{k}}\;
{\cal R}^{c_3 d_3}_{\;\;\;\;\;\;a_3b_3}\cdots {\cal S}^{c_j d_j}_{\;\;\;\;\;\;a_j d_j}\cdots
{\cal R}^{c_{k} d_{k}}_{\;\;\;\;\;\;a_{k}b_{k}}\;
\label{vary2}
\ee
and
\be
\frac{1}{k}\frac{\delta^2 \left({{\cal S}}\cdot
{\cal X}_k\right)}{\delta{\cal S}_{ef}^{\;\;\;\;gh}\partial{\cal R}_{ab}^{\;\;\;\;cd}} \;=\; (k-1)
\;\delta^{ab  a_2b_2\ldots ef \ldots  a_{k}b_{k}}_{cd c_2d_2 \ldots gh \ldots
 c_{k} d_{k}}
\;{\cal R}^{c_2 d_2}_{\;\;\;\;\;\;a_2b_2}\cdots[\ldots]\cdots
{\cal R}^{c_{k} d_{k}}_{\;\;\;\;\;\;a_{k}b_{k}}\;,
\label{vary1}
\ee
where, in the last line, the symbol
$[\ldots]$ indicates the absence of ${\cal S}^{c_j d_j}_{\;\;\;\;\;\;a_jb_j}$
and
 $\;e,f,g,h\;$ are standing in place of $\;a_j,b_j,c_j,d_j\;$
respectively.

From Eqs.~(\ref{vary2}) and~(\ref{vary1}), it follows
that both terms on the left side of  Eq.~(\ref{rerun})
vanish by virtue of the second Bianchi identity with respect to  permutations
of $a$, $e$ and $f$.

In summary, we have shown that any of the Lovelock higher-derivative gravity theories has an effective description as Einstein gravity non-minimally coupled to a ($D-2$)-form field-strength tensor. So that, just as an $f(R)$ theory is Einstein
gravity coupled to a scalar, any higher-derivative  unitary theory of gravity is Einstein gravity coupled to a ($D-3$)-form gauge field.  The implication is that, for all practical purposes, Einstein's is the single unitary theory of gravity.
Our constructions would fit in naturally  with the myriad of string-theory models that include such higher-form gauge fields.

\section*{Acknowledgments}

The research of RB was supported by the Israel Science Foundation grant no. 239/10.
The research of AJMM  was  supported by  a  Rhodes University Discretionary Grant RD11/2012. AJMM thanks Ben Gurion University for their hospitality during his visit.

\end{document}